\documentclass[superscriptaddress,groupedaddress,nofootnoteinbib,11pt]{article}
\pdfoutput=1 
\usepackage{jheppub}
\usepackage{CJKutf8} 
\usepackage{mathtools,slashed,mathrsfs}
\usepackage[caption=false]{subfig}
\usepackage{dcolumn}
\usepackage{multirow}
\usepackage{tabularx}
\usepackage{booktabs}
\usepackage{bm}
\usepackage{breqn}
\usepackage{comment}
\usepackage{setspace}
\usepackage[dvipsnames]{xcolor}
\usepackage[normalem]{ulem} 
\usepackage{enumerate}
\usepackage{siunitx}

\definecolor{mypurple}{RGB}{164,64,214}

\newcommand\nn{\nonumber}
\newcommand\eea{\end{eqnarray}}
\newcommand\bea{\begin{eqnarray}}
\newcommand\ees{\end{split}}
\newcommand\bes{\begin{split}}

\def\l{\left(}
\def\r{\right)}

\def\rarrow{\rightarrow}
\def\ket{\right \rangle}
\def\bra{\left\langle}
\newcommand{\mA}{m_A}
\newcommand{\Ad}{A_d}
\newcommand{\Adsigma}{A_{d \, \sigma}}
\newcommand{\Admu}{A_{d \, \mu }}
\newcommand{\vecAd}{\Vec{A}_{d}}
\newcommand{\Deltanu}{\Delta_{\nu}}
\newcommand{\gammasigma}{\gamma^{\sigma}}

\begin{document}

\title{Constraining Vector Dark Matter with Neutrino experiments}

\author[a]{Dawid Brzeminski,}
\author[a]{Saurav Das,}
\author[a]{Anson Hook,}
\author[a]{and Clayton Ristow}

\affiliation[a]{Maryland Center for Fundamental Physics, University of Maryland, College Park, MD 20742}

\emailAdd{dbrzemin@umd.edu}
\emailAdd{sauutsab@umd.edu}
\emailAdd{hook@umd.edu}
\emailAdd{cristow@umd.edu}

\abstract{
Vector Dark Matter (VDM) that couples to lepton flavor ($L_e$, $L_{\mu}$, $L_{\tau}$) acts similarly to a chemical potential for the neutrino flavor eigenstates and modifies neutrino oscillations.  VDM imparts unique signatures such as time and directional dependence with longer baselines giving better sensitivity. 
We use the non-observation of such a signal at Super-Kamiokande to rule out the existence of VDM in a region of parameter space several orders of magnitude beyond other constraints and show the projected reach of future experiments such as DUNE.
}
\maketitle

\section{Introduction}

Determining the microscopic properties of dark matter is among the greatest puzzles in modern physics.  There is overwhelming evidence~\cite{Bergstrom:2000pn,Bertone:2004pz,Lisanti:2016jxe} for the existence of dark matter (DM) but many of its basic properties, such as its mass, spin and interactions remain completely unknown.

In this article, we will focus on Ultralight Dark Matter (ULDM). ULDM are well-motivated DM candidates~\cite{Arvanitaki:2009fg,Hui:2016ltb} with rich experimental signatures.  Due to their very low mass, the number of particles per Compton volume is large, $n_{DM} \lambda_{\text{Compton}}^{3} \gg 1$, which allows us to treat them as a classical field rather than as a set of particles.  
As the local DM field has an average velocity of $v\sim 10^{-3}$, the field's energy dispersion is very small, $\Delta E \sim m v^2 = 10^{-6} m$ and the field changes over very long length scales $l \sim 1/p \sim 10^{-3}/m$. As a result, we can approximate the field as coherently oscillating with the frequency $\omega = m $ until timescales of order $t_{\text{coh}} \sim (m v^2)^{-1} = 10^{-6} m^{-1}$ before the frequency spread becomes important. 
The long coherence length and time of dark matter lead to unique signatures.  When the dark matter mass is so small that the long coherence length wipes out structure on observably small length scales, ``fuzzy dark matter" phenomenology kicks in~\cite{Hu:2000ke,Irsic:2017yje,Dentler:2021zij,Hui:2021tkt}.
Additionally, the oscillation in time of the dark matter field can lead to observable time-dependent effects if DM couples to the Standard Model (SM)
~\cite{Khmelnitsky:2013lxt,Graham:2015ouw,Graham:2015ifn,Arvanitaki:2014faa,Geraci:2018fax,Irastorza:2018dyq, PhysRevA.93.063630}. For instance, ULDM QCD axions~\cite{Hook:2018jle,DiLuzio:2021gos,DiLuzio:2021pxd} result in the electric dipole moment of the neutron varying with the frequency of the axion mass~\cite{Graham:2013gfa,Budker:2013hfa}.

In this work, we focus on a spin one ULDM ($\vecAd$), for related work see Refs.~\cite{ADMX:2010ubl,Chaudhuri:2014dla,Knirck:2018ojz,An:2014twa,Bloch:2016sjj}.  
The main difference between scalars and vectors is polarization.  Much like the amplitude of the field, the polarization changes direction on timescales of order $t_{\text{coh}}$.  For sufficiently long DM coherence times, earth-based experiments sensitive to the direction of the polarization would see the polarization of dark matter rotating as the Earth spins around its axis.  This would introduce a daily modulated feature on top of the standard oscillation with the frequency of the dark matter mass, which is a signature that can be used to distinguish scalar from vector DM.  Additionally, experiments with angular resolution would see strong directional dependence. 

Neutrino oscillations are uniquely suited for exploring dark matter interactions.  As long as the interactions are flavor non-universal, neutrino oscillations are sensitive to energy differences of the order $\Delta m_\nu^2/E \sim 10^{-12}$ eV, over 15 orders of magnitude more sensitive than traditional dark matter detectors.  The requirement that the interaction is flavor non-universal combined with our focus on VDM immediately leads us to consider $U(1)_{L_{\mu}-L_{\tau}}$ and $U(1)_{L_{e}-L_{\mu}}$.

New vectors interacting with the SM are extremely common. Kinetic mixing with the SM photon, gauging $U(1)_{B-L}$ or $U(1)_{L_{\mu}-L_{\tau}}$ are among the most commonly considered scenarios~\cite{Carney:2019cio,Pierce:2018xmy,Dror:2019uea,Fabbrichesi:2020wbt}. The $U(1)_{L_{\mu}-L_{\tau}}$ and $U(1)_{L_{e}-L_{\mu}}$ gauge bosons that we consider have a plethora of interesting implications for neutrino physics~\cite{Dev:2020kgz,Reynoso:2016hjr,Berlin:2016woy, Krnjaic:2017zlz, Brdar:2017kbt, Davoudiasl:2018hjw,Liao:2018byh,Capozzi:2018bps,Huang:2018cwo,Farzan:2019yvo,Cline:2019seo,Losada:2021bxx,Huang:2021kam,Chun:2021ief,Losada:2022uvr,Dev:2022bae}.

A background of $U(1)_{L_{\mu}-L_{\tau}}$ ($U(1)_{L_{e}-L_{\mu}}$) gauge bosons acts as a time-dependent potential that is flavor non-universal.  
As neutrinos and antineutrinos have opposite charges, the potential is opposite for matter and antimatter.  Additionally, the effect is proportional to the dot product of the neutrino velocity with the polarization of dark matter resulting in a directionally dependent chemical potential.  
If the dark matter oscillations are slow compared to the neutrino propagation timescale, the effect on neutrino oscillations is similar to the MSW matter effect \cite{Wolfenstein:1977ue,Mikheyev:1985zog,Super-Kamiokande:2017yvm}. Therefore, the main signature of these models are time-dependent neutrino oscillation parameters, e.g. $\Delta m_{31}^{2}(t)$ and $\sin \theta_{13}(t)$. Measurement of these signatures might be challenging due to a low neutrino event rate. However, we show that even after averaging over the duration of the experiment, the effect is strong enough to place bounds several orders of magnitude better than the leading bounds on $U(1)_{L_{e}-L_{\mu}}$ and $U(1)_{L_{\mu}-L_{\tau}}$ dark matter. Similar bounds on $U(1)_{L_{\mu}-L_{\tau}}$ from neutrino oscillation experiments were estimated in \cite{Brdar:2017kbt} and \cite{Alonso-Alvarez:2021pgy}. The former studied the effect of neutrino-DM scattering, while the latter considered the scenario where the vector field is a relic field used to facilitate the production of sterile neutrino dark matter in the early universe.

If the dark matter oscillations are fast compared to the neutrino propagation timescale, the neutrinos experience a rapidly oscillating effective chemical potential. Because this potential oscillates rapidly, to the lowest order the effective potential averages to zero over the neutrino propagation time. Thus, the leading effects are suppressed by an additional factor of $m^{-1}$, where $m$ is the oscillation rate of the background dark matter field. The fact that our bounds on $U(1)_{L_{e}-L_{\mu}}$ and $U(1)_{L_{\mu}-L_{\tau}}$ scale by an additional factor of $m^{-1}$ in the high mass regime reflects this chemical potential washout effect.

The goal of this paper is to explore the sensitivity of various neutrino oscillation experiments to $U(1)_{L_{\mu}-L_{\tau}}$ ($U(1)_{L_{e}-L_{\mu}}$) DM. In section~\ref{Sec: Model}, we review the phenomenology of neutrino oscillations and introduce the effect of vector DM on neutrino oscillations. In section~\ref{Sec: Experiments}, we constrain the model using existing and planned neutrino experiments.  We conclude with section~\ref{Sec: Conclusions} and propose further avenues of research.

\section{Model}
\label{Sec: Model}

We start with a brief description of gauging lepton flavor and its effect on neutrinos. 
The three individual lepton flavor numbers $L_{\alpha},\ \alpha=e,\mu,\tau$,  can be equivalently described in a different basis comprising of the total lepton number $\sum_{\alpha} L_{\alpha}=L$ and two lepton number differences, which we choose to be $L_e-L_{\mu}$ and $L_{\mu}-L_{\tau}$ without any loss of generality. The total lepton number charge $L$ is the identity in the flavor space and gives an equal phase to neutrinos of all flavors and hence is unobservable in neutrino oscillations. Therefore we restrict our discussion to $L_e-L_{\mu}$ and $L_{\mu}-L_{\tau}$.
 
To the Standard Model, we add a single gauge boson $\Admu$ which gauges either of the two flavor lepton number differences. 
\bea
\label{Eq:Lagrangian}
&&\mathcal{L}\subset-\frac{1}{4}F_{d\, \mu\nu }F_d^{\mu\nu}-\frac{1}{2}\mA^2\Admu\Ad^{\mu}+\mathcal{L}_{int,e-\mu} +\mathcal{L}_{int,\mu-\tau}\nn \\
&& \mathcal{L}_{int,e-\mu}=-g \Adsigma \l  \overline{l_e}\gammasigma l_e-\overline{l_{\mu}}\gammasigma l_{\mu}\r \\
&&\mathcal{L}_{int,\mu-\tau}=-g' \Adsigma \l\overline{l_{\mu}}\gammasigma l_{\mu}-\overline{l_{\tau}}\gammasigma l_{\tau}\r\nn
\eea

where  $\mA$ is the mass of the gauge field, $g$ and $g'$ denote coupling to the $L_e-L_{\mu}$ and $L_{\mu}-L_{\tau}$ charges respectively and $l_{e,\mu,\tau}$ denotes leptons with $e,\mu,\tau$ charges. For example, $l_e$ denotes electrons, electron neutrinos and their anti-particles.

We now consider the consequences of the vector boson $\Admu$ being dark matter. The three polarizations of a massive gauge field satisfy $p_{\mu}\epsilon^{\mu}=0$.  If the gauge bosons were at rest, the three polarizations would simply be unit vectors in the x, y, and z directions.  Including a non-zero velocity along the z-axis and considering the longitudinal polarization $\epsilon_L=\frac{1}{\mA}(p, 0,0,E)$ we see that the temporal component of the vector field $A_{d\, 0}$ is subdominant to the spatial components $A_{d\,i}$ by a factor of DM velocity $v_{DM}\sim 10^{-3}$. To leading order the galactic DM field can be written as an oscillating three vector with frequency $\omega\sim \mA+ \mathcal{O}(v^2_{DM})$
\bea\label{Eq:DMbackground}
\vecAd=\Vec{A}_{0}\cos\l\mA (t+\vec{v}_{DM}\cdot \vec{x})+\phi\right)+\mathcal{O}(v_{DM}^2) ,
\eea
where $\vert \Vec{A}_{0} \vert =  \sqrt{2\rho_{DM}/\mA^2}$ and $\mA$ is the mass of the DM. 
Since the neutrinos are relativistic, the interactions in Eq.~\ref{Eq:Lagrangian} can be written for neutrinos as
\bea\label{Eq:dotproduct}
H_{\text{int}}=\vecAd\cdot \Vec{v}_{\nu}\begin{pmatrix}
g & 0 & 0 \\
0 & g'-g & 0 \\
0 & 0 & - g'
\end{pmatrix} +\mathcal{O}(v_{DM})
\eea
where $\Vec{v}_{\nu}$ is the neutrino velocity. The dark matter shifts the relative energies of different neutrino flavor eigenstates. For example, in the case where the dark matter is the gauge boson of $L_e-L_{\mu}$, the dark matter creates an energy difference between the electron neutrino $\nu_e$ and muon neutrino $\nu_{\mu}$. The vector nature of the interaction means it has opposite signs for neutrinos and anti-neutrinos. The dark matter thus acts as a chemical potential for the neutrinos. This can be contrasted with the effect of a scalar field which does not distinguish between particles and antiparticles.  

It is important to note the directional dependence of the interaction. The interaction picks the polarization of the DM along the neutrino velocity. In neutrino beam experiments, where the neutrino velocity has a fixed orientation with respect to the earth's frame, the direction of the neutrino velocity rotates with the earth over the course of a day. This produces a daily modulation of the DM effect. This is in contrast with the effect produced by a scalar field where no such directional dependence exists and can be used to distinguish between the two. We explore this effect in more detail in Sec. \ref{Subsec: Daily Modulation}

\subsection{Neutrino Oscillations}

Within the Standard Model, neutrino oscillations are controlled by two contributions, the vacuum Hamiltonian $H_{\text{vac}}$  and the MSW effect \cite{Wolfenstein:1977ue, Mikheyev:1985zog}. In absence of any matter, the neutrino oscillation is set by the neutrino energy, the mass squared difference and the rotation matrix between the mass and the flavor eigenstates, $U_{\text{PMNS}}$. In the relativistic limit, we can expand the Hamiltonian in powers of $m_{\nu}/E_{\nu}$ where $m_{\nu}$ denotes the mass of a generic mass eigenstate and $E_{\nu}$ is the neutrino energy. The non-identity piece of the vacuum Hamiltonian can be written as 
\bea
H_{\text{vac}} = U_{PMNS} 
\left[ \frac{1}{2E_{\nu}} \begin{pmatrix} 
0 & 0 & 0 \\
0 & \Delta m_{21}^{2} & 0 \\
0 & 0 & \Delta m_{31}^{2}
\end{pmatrix} \right] U_{PMNS}^{\dagger}
\eea
where $\Delta m^2_{ij}=m^2_i-m^2_j$ are the neutrino mass differences and we have removed the pieces proportional to the identity that do not contribute to oscillations.

On the other hand, in the presence of matter, for example while propagating through the earth, the interaction of the neutrinos with matter creates a chemical potential for the neutrinos. In the presence of protons, neutrons and electrons, the only non-identity interaction in flavor space is a chemical potential for the electron neutrinos.
\bea \label{Eq: MSW hamiltonian}
H_{\text{MSW}} =  \pm \sqrt{2} G_{F} n_{e}(x)
\begin{pmatrix}
1 & 0 & 0 \\
0 & 0 & 0 \\
0 & 0 & 0
\end{pmatrix}
\eea
where $G_F$ is the Fermi constant, $n_e(x)$ is the number density of electrons and  the $+$ ($-$) sign is for neutrinos (anti-neutrinos).
The neutrino oscillations we observe in neutrino experiments are an interplay between the vacuum oscillation and the effect of matter. Experiments that are sensitive to this effect work in the regime $L \Delta m_{21}^{2}/E_{\nu} \ll 1$, despite the long distance traveled by neutrinos $L\sim 10^3-10^4$ km. As a result, the relevant mixing parameters are reduced to $\Delta m_{31}^{2}$, $\theta_{13}$ and $\theta_{23}$. When there is a non-zero number density of electrons, $\Delta m_{31}^{2}$ and $\theta_{13}$ are replaced by their effective counterparts \cite{Super-Kamiokande:2017yvm}

\bea
\Delta m^{2}_{31,M} = \Delta m^{2}_{31} \sqrt{(\cos 2\theta_{13} - \Gamma)^{2} + \sin^{2} 2 \theta_{13}}\\
\sin^{2} 2\theta_{13,M} = \frac{\sin^{2} 2\theta_{13}}{(\cos 2\theta_{13} - \Gamma)^{2} + \sin^{2} 2 \theta_{13}}
\eea
where $\Gamma = \sqrt{2} G_{F} n_e E_{\nu} / \Delta m_{31}^{2}$. 
When $\Gamma = \cos 2\theta_{13}$, magnitudes of $H_{\text{vac}}$ and $H_{\text{MSW}}$ are comparable, which significantly changes mass eigenstates. This leads to enhanced oscillations into electron neutrinos at distances of order $\frac{E \nu}{\Delta m_{31,M}^2}$, which we call resonance. 

The effect has important phenomenological consequences. The sign of $\Gamma$ changes between neutrinos and antineutrinos, and also depends on mass ordering. In the normal ordering, the resonance can only happen for neutrinos, while in the inverted ordering it can only occur for antineutrinos. Therefore, the resonance can be used to solve the mass ordering problem \cite{Super-Kamiokande:2017yvm}.

In the presence of vector dark matter, neutrino oscillations are modified in a similar way. The dark matter contribution (Eq.~\ref{Eq:Lagrangian}) to the neutrino oscillation can be written as\footnote{Although the DM phase changes both in space and time, as in Eq.~\ref{Eq:DMbackground}, the neutrino baseline is much smaller than the coherence length so that the change in phase is suppressed by the small DM velocity. Hence we only keep the leading time-dependent piece.} 
\bea\label{Eq:intHamiltonian}
H_{\text{int}} &=&\pm  \frac{\sqrt{2 \rho}}{\mA} \cos(\mA t+ \phi) \cos{\alpha(t)}
\begin{pmatrix}
g & 0 & 0 \\
0 & g'-g & 0 \\
0 & 0 & - g'
\end{pmatrix} 
\eea
where $g$ and $g'$ denote coupling to the $L_e-L_{\mu}$ and $L_{\mu}-L_{\tau}$ charges respectively,  $\rho$ is the DM energy density, $\mA$ is the mass of the vector DM and $\alpha(t)$ is the angle between DM polarization and the neutrino velocity. The total Hamiltonian governing neutrino oscillations is the sum of the parts
\bea
H = H_{\text{vac}} + H_{\text{MSW}} + H_{\text{int}} .
\eea

In order to see the effect of the VDM is similar to the matter effect, let us set $g = 2g'$. Then, the non-identity piece of $H_{\text{int}}$ has the same matrix structure as Eq. \eqref{Eq: MSW hamiltonian}. In the $L \Delta m_{21}^{2}/E_{\nu} \ll 1$ limit this leads to a time-dependence of the effective $\Delta m_{31}^{2}$ and $\theta_{13}$ 
\bea
\Delta m^{2}_{31,DM} = \Delta m^{2}_{31} \sqrt{(\cos 2\theta_{13} - \Gamma - \Gamma_{DM}(t))^{2} + \sin^{2} 2 \theta_{13}}\\
\sin^{2} 2\theta_{13,DM} = \frac{\sin^{2} 2\theta_{13}}{(\cos 2\theta_{13} - \Gamma -\Gamma_{DM}(t))^{2} + \sin^{2} 2 \theta_{13}}
\eea
where $\Gamma_{DM} (t) = \frac{3 g \sqrt{ \rho} E_{\nu}}{\sqrt{2} m \Delta m_{31}^{2}}   \cos(m t+ \phi) \cos \alpha(t) $. Therefore, we can see that experiments which are sensitive to the matter effect should also be able to measure the effect caused by DM. The only qualitative difference is that in this scenario the time-averaged effect is the same for neutrinos and anti-neutrinos, which is a feature that could be used to differentiate the DM-induced effect from the matter effect.

\subsection{Simplified model with two flavors}

To gain some insight into the effect of vector dark matter on neutrino oscillation, we consider a toy model of two neutrino flavors. For clarity, we ignore the matter effect as it does not affect our conclusions.  The model is characterized by a single neutrino mass squared difference, which we denote by $\Delta m^2$, the neutrino energy $E_{\nu}$ and the dark matter mass and coupling, $\mA$ and $g$.
\bea
        H&&=U\frac{\Delta m^2}{4 E_{\nu}}\sigma_z U^{\dagger}+\frac{g\sqrt{2\rho}}{\mA}\sigma_z\cos(\mA (t -t_0) +\phi(t_0))+\mathcal{O}(v_{\text{DM}})\nn\\
        &&=U\Deltanu\sigma_z U^{\dagger}+g A_0\sigma_z \cos(\mA (t -t_0) +\phi(t_0)) \, ,
\eea
where $\sigma_z$ is the third Pauli Matrix, $U$ is the PMNS matrix for two flavors, $\Deltanu\equiv \frac{\Delta m^2}{4 E_{\nu}}$ controls the neutrino vacuum Hamiltonian, $A_0=\frac{\sqrt{2\rho}}{\mA}$ is the vector amplitude, $\phi(t_0)=\mA t_0+\phi_0$  is the DM phase at the point of neutrino production and $t_0$ denotes the time of neutrino production.
We have also neglected the dot product of the polarization and the neutrino velocity which changes due to earth's rotation.  We discuss this effect in detail in Sec.~\ref{Subsec: Daily Modulation}.

Let us first consider the time dependence of the DM effect. Due to stringent constraints, the DM effect is subdominant to the vacuum oscillation and can be treated using perturbation theory. The transition probability between the flavor states can be written as
\bea\label{Eq:P0P2}
P_{\nu_e\rarrow\nu_{\mu}}=P^{(0)}+ P^{(1)}+P^{(2)}+\cdots\, ,
\eea
where we have labeled the two flavor states as $\nu_e$ and $\nu_{\mu}$, $P^{(1)}$ and $P^{(2)}$ are the first and second order correction to the oscillation probability due to dark matter, $P^{(n)}\propto g^n$.
The first order correction, $P^{(1)}$, only contains contributions as $\sin(\mA L+\phi(t_0))$, where $L$ denotes the baseline of the experiment. This term oscillates as the initial phase $\phi (t_0)$ changes as a function of neutrino production time $t_0$. If an experiment has sufficient statistics, they can look for the oscillating first-order correction.
On the other hand, if the experiment only observes time-averaged quantities, we will instead be sensitive to the second order correction, $P^{(2)}$, which includes terms like $\sin^2(\mA L+\phi(t_0))$. These do not average to zero in contrast with the first-order correction. Then,
\bea\label{Eq:phaseaverage}
\langle P_{\nu_e\rarrow\nu_{\mu}}\rangle=P^{(0)}+\langle P^{(2)}\rangle+\cdots \, ,
\eea
where $\langle \rangle$ denotes averaging over the lifetime of the experiment. We compute $\langle P^{(2)}\rangle$ using time-dependent perturbation theory.

\bea\label{Eq:smallbaseline}
&&\langle P^{(2)}\rangle=-\frac{g^2A_0^2\sin ^2(2 \theta )}{ \l\mA^3-4 \mA \Delta _{\nu }^2\r^2} f(\mA,\Deltanu,L)\nn \\
&&f(\mA,\Deltanu,L)=8 \Delta _{\nu }^4 \cos ^2(2 \theta ) \left(\cos \left(L \left(m_A+2 \Delta _{\nu }\right)\right)+\cos \left(L \left(m_A-2 \Delta _{\nu }\right)\right)-2 \cos \left(2 L \Delta _{\nu }\right)\right)\nn \\
&&+m_A^4 \left(L \Delta _{\nu } \sin ^2(2 \theta ) \sin \left(2 L \Delta _{\nu }\right)+2 \cos ^2(2 \theta ) \sin ^2\left(L \Delta _{\nu }\right)\right)+16 m_A \Delta _{\nu }^3 \cos ^2(2 \theta ) \sin \left(L m_A\right) \sin \left(2 L \Delta _{\nu }\right)\nn\\
&&+m_A^2 \Delta _{\nu }^2 \left(\cos (4 \theta ) \left(\cos \left(L \left(m_A+2 \Delta _{\nu }\right)\right)+\cos \left(L \left(m_A-2 \Delta _{\nu }\right)\right)-2 \cos \left(L m_A\right)+6 \cos \left(2 L \Delta _{\nu }\right)-6\right)\right)\nn\\
&&+m_A^2 \Delta _{\nu }^2 \left(\cos \left(L \left(m_A+2 \Delta _{\nu }\right)\right)+\cos \left(L \left(m_A-2 \Delta _{\nu }\right)\right)+2 \cos \left(L m_A\right)\right)\nn\\
&&+m_A^2 \Delta _{\nu }^2 \left(-4 L \Delta _{\nu } \sin ^2(2 \theta ) \sin \left(2 L \Delta _{\nu }\right)+2 \cos \left(2 L \Delta _{\nu }\right)-6\right) ,
\eea 
where $\theta$ characterizes the two flavor PNMS matrix $U$. 

This very complicated expression can be greatly simplified in the large or small $\mA$ limits. In the small mass limit, where $\mA \ll 1/L\sim \Delta_\nu$,  the effect of dark matter is roughly constant as the neutrinos fly from the point of production to the point of detection.
In this limit, the leading correction to the vacuum oscillation is given by
\bea\label{Eq:lowmass}
&&\langle P^{(2)}\rangle = g^2 \frac{A_0^2\sin^2 2\theta}{8\Deltanu^2}  \\
&& \l 4 \cos 4 \theta +2 \cos (2 \Deltanu  L) \l\cos 4 \theta(\Deltanu ^2 L^2-2)+\Deltanu ^2 L^2-1\r -\Deltanu  L (5 \cos (4 \theta )+3) \sin (2 \Deltanu  L)+2\r\nn.
\eea
On the other hand in the large mass limit, where $\mA \gg 1/L\sim \Delta_\nu$, the dark matter is oscillating rapidly over the flight of the neutrinos and much of its effect is averaged away. In this limit, the same second-order correction reads

\bea\label{Eq:highmass}
\langle P^{(2)}\rangle=-g^2\frac{A_0^2 \sin ^22 \theta }{\mA^2} \l\Deltanu L \sin ^2(2 \theta ) \sin (2 \Deltanu  L)+2 \cos ^2(2 \theta ) \sin ^2(\Deltanu  L)\r 
\eea

\begin{figure}[t]
    \centering
    \includegraphics[width=0.65\textwidth]{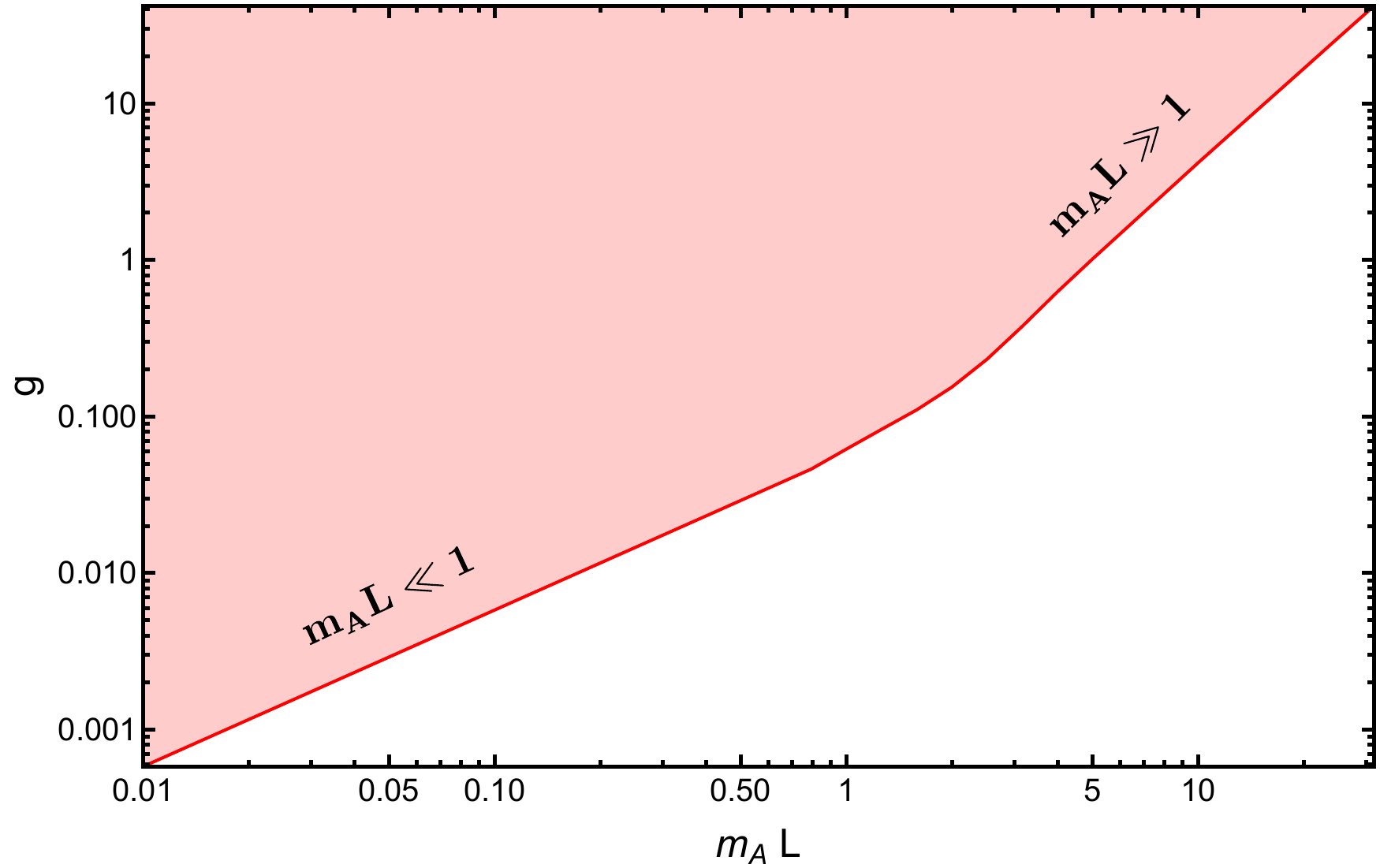}
    \caption{We show how the bound on the dark matter gauge coupling as a function of mass is expected to behave for hypothetical neutrino oscillation experiments. For $\mA L\ll 1$, for a fixed experimental uncertainty, the coupling scales linearly with $\mA$ whereas for $\mA L\gg 1$, the same scaling is quadratic.}
    \label{fig: perturbation}
\end{figure}

From Eq.~\ref{Eq:lowmass} and ~\ref{Eq:highmass}, it is easy to see how the bound on the dark matter coupling constant $g$ scales with the mass of dark matter.  In the low mass regime, $\sqrt{P^{(2)}} \sim g A_0\sim\frac{g}{\mA}$, so we expect the constraint to scale as $g \propto \mA$. On the other hand, in the high mass regime, $\sqrt{P^{(2)}} \sim \frac{gA_0}{\mA}\sim\frac{g}{\mA^2}$.
Hence the constraint is expected to scale as $g \propto \mA^2$. 
This scaling behavior can be seen in  Fig.~\ref{fig: perturbation} where we have plotted the bound obtained from a hypothetical neutrino experiment using two neutrino flavors in the perturbative regime.  In making this plot, we have assumed $\Deltanu\in[1.,2.]$, a fixed experimental uncertainty and considered 10 bins with equal energy spacing (motivated by the DUNE experiment~\cite{DUNE:2015lol,DUNE:2020fgq,DUNE:2020jqi}).  The difference in the scaling of the bound on the coupling constant is clearly different in the small and large mass ranges.  In the low mass regime, $\mA L\ll 1$, the bound increases slowly $g\propto \mA$. On the other hand, in the high mass regime $\mA L\gg 1$ it weakens faster, $g\propto \mA^2$ and smoothly transitions between the two at $\mA L\sim 1$.

In Section ~\ref{Sec: Results}, we will use these perturbative results to scale numerically obtained bounds at one 
mass to other masses.  To do this, we will need to be sure that we are in the perturbative regime.  We will look at
experiments with neutrino energies $E_\nu\sim  \mathcal{O} (1-10 \, \text{GeV})$ at which scale $\Delta m_{13}^2$ 
dominates the oscillation dynamics. By demanding $P^{(2)}\ll 1$, Eq.~\ref{Eq:lowmass} gives a bound on the coupling below which we are working in the perturbative regime. 
\bea
\label{Eq: Perturbative limit}
g<2.5\cdot 10^{-30} \l \frac{\mA}{10^{-20} \text{eV}}\r  \l \frac{\text{GeV}}{E_\nu}\r
\eea

\subsection{Daily Modulation}\label{Subsec: Daily Modulation}

In this section, we discuss the effects of the earth's rotation on our DM signal. In particular, we show that the effect of daily modulation does not erase the DM signal even after time-averaging, which motivates the search for time-independent signatures of the signal.

As the DM-induced potential is the dot product between the DM polarization and neutrino velocity, the DM-induced potential undergoes an oscillation due to the earth's rotation.
The equatorial component oscillates whereas the polar components remain unchanged due to the earth's rotation. 
This directional dependence is a crucial difference between vector and scalar DM.

\begin{figure}[t]
    \centering
    \includegraphics[width=0.75\textwidth]{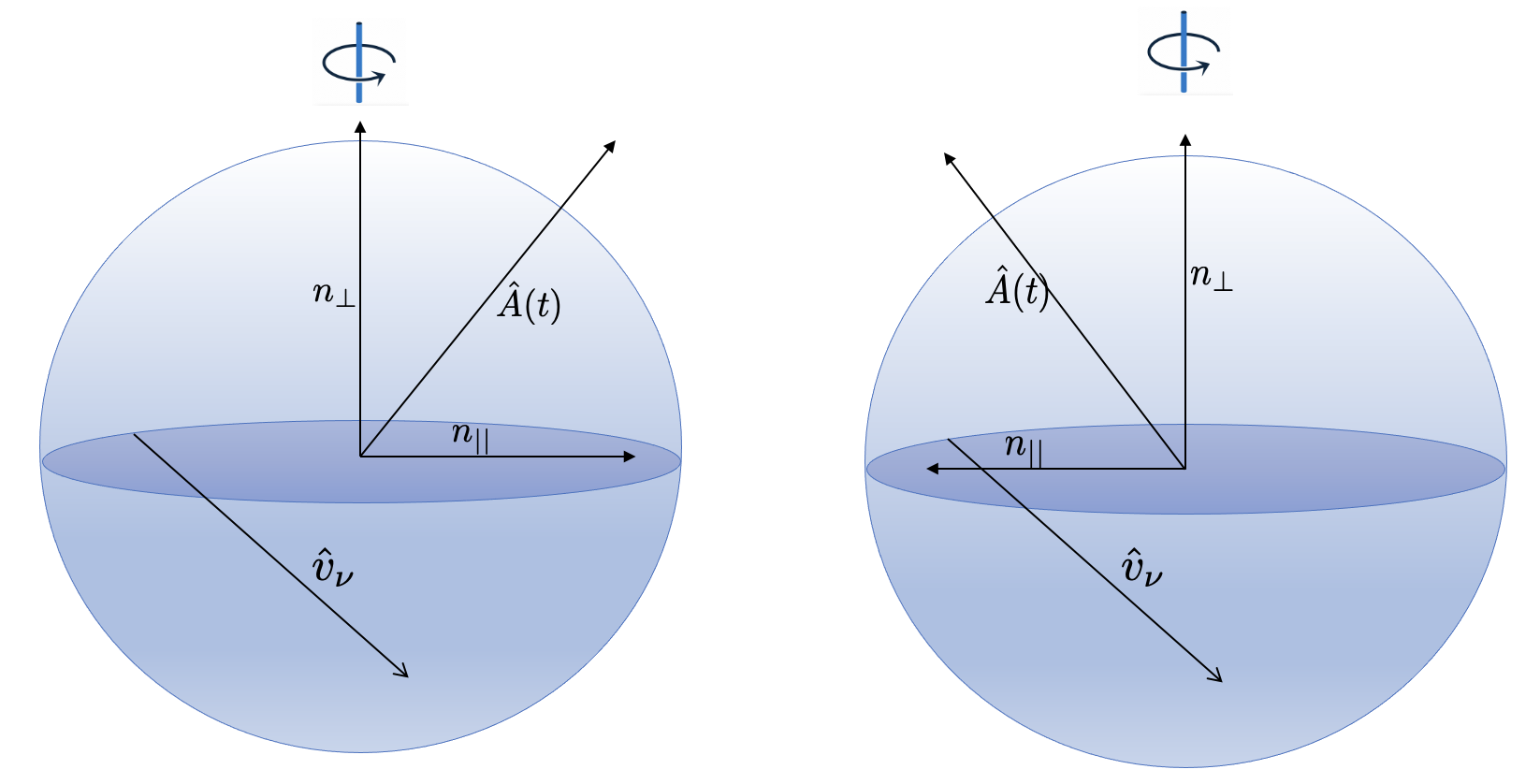}
    \caption{ The equatorial component of the vector polarization $n_{||}$ rotates in the lab frame due to the earth's rotation whereas the polar component $n_{\perp}$ remains unchanged. For experiments like DUNE, the neutrino velocity has a fixed direction. Hence the angle between the vector polarization and the neutrino velocity oscillates with the frequency of a day. }
    \label{fig:DUNE_earth_rotation}
\end{figure}

For a neutrino beam type experiment, the neutrino beam has a fixed direction in the lab frame. Figure~\ref{fig:DUNE_earth_rotation} shows this in the lab frame.
In general, the neutrino beam has a direction
\bea
\hat{v}_{\nu}=(v_{||},\ \ v_{\perp})
\eea
where $v_{||}$ and $v_{\perp}$ are the equatorial and polar components of the neutrino velocity respectively. We can similarly write the dark matter polarization unit vector in the lab frame as
\bea
\hat{A}(t)=(n_{||}\cos(\omega_d t),\ \ n_{\perp}) ,
\eea
where $n_{||}$ and $n_{\perp}$ are the equatorial and polar components of the vector polarization and $\omega_d$ is the daily frequency. We can write the vector DM potential, Eq.~\ref{Eq:intHamiltonian},
\bea
&&H_{\text{int}} =  \frac{\sqrt{2 \rho}}{\mA} \cos(\mA t+ \phi)\cos\alpha(t)
\begin{pmatrix}
g & 0 & 0 \\
0 & g'-g & 0 \\
0 & 0 & - g'
\end{pmatrix} \\
&&\cos\alpha(t) = \hat{A}(t)\cdot\hat{v}_{\nu} .
\eea
In an experiment with high statistics, we can make out the daily modulation of the signal. In absence of such opportunity, we can observe the time averaged effect of the daily modulation.

The daily average removes the first order correction $\langle P^{(1)}\rangle=0$ and only the higher order effects remain observable. In the case of sufficiently light DM, $\mA\ll\omega_d$, averaging over a day does not average over a full oscillation of DM.  
In that case, the oscillation of the DM effect is observable even after taking the daily average.  On the other hand, if one averages over the lifetime of the experiment, as is typically done, the dark matter oscillations are also averaged over.  

After averaging over the lifetime of the experiment, the oscillation probability is given by
\bea\label{Eq:daily average}
\langle P_{\nu_e\rarrow\nu_{\mu}}\rangle
&=&P^{(0)}+\l\frac{1}{2}n_
{||}^2v_{||}^2+n_{\perp}^2v_{\perp}^2\r  \langle P^{(2)}\rangle+\cdots
\eea
where $P^{(0)},P^{(2)}$ is same as in Eq.~\ref{Eq:P0P2}. Here we have assumed that the DM is light enough so that the coherence time of the DM is longer than the experiment lifetime, hence the vector  polarization ($n_
{||},n_{\perp} $) can be treated as constant. We see that the effect of the daily average is a correction of the DM effect which is dependent on the DM polarization and the beam direction. Given a global network of neutrino beam experiments, we can not only observe but also map out the DM polarization in the solar neighborhood.

\begin{figure}[h]
    \centering
    \includegraphics[width=0.75\textwidth]{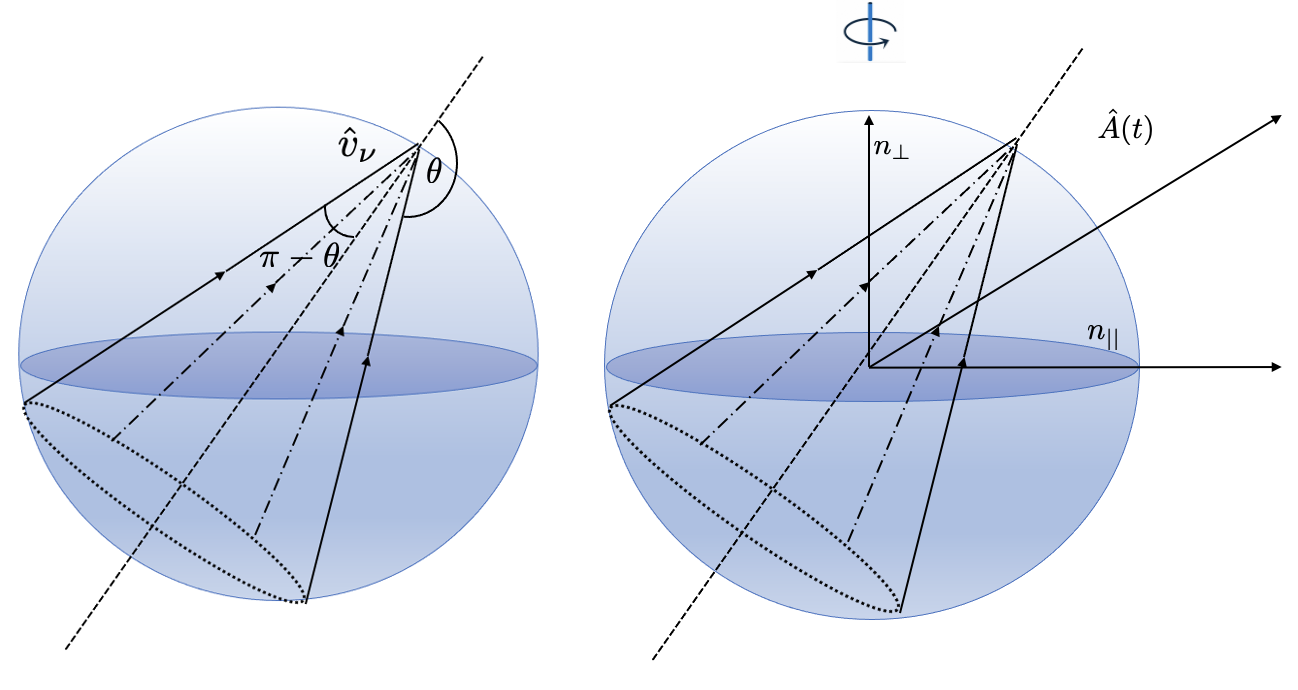}
    \caption{Schematic diagram showing the two different averages over atmospheric neutrino velocity. For atmospheric neutrinos, the neutrino velocity does not have a fixed direction. On the left, we have shown the averaging of the neutrino velocity over the same zenith angle $\theta$ (it is conventional to assign $\theta=0$ to down-going neutrinos). Paths with the same zenith angle have equal lengths which form a cone. This is standard for atmospheric neutrino experiments, as discussed in \ref{subsec:SuperK}. We also have to average over the daily modulation, which we have shown on the right.}
    \label{Fig:Atmosphericaverage}
\end{figure}

If the DM coherence time is longer than a day but much smaller than the lifetime of the experiment, we have to average over all possible DM polarization. 
\bea\label{Eq:angularaverage}
\langle P_{\nu_e\rarrow\nu_{\mu}}\rangle
&=&P^{(0)}+\frac{1}{2}\l\frac{1}{2}v_{||}^2+v_{\perp}^2\r \langle P^{(2)}\rangle +\cdots
\eea

For a neutrino beam experiment like DUNE, we can compare Eq.~\ref{Eq:phaseaverage} to Eq.~\ref{Eq:angularaverage} to perceive the effect of the earth's rotation. The difference depends upon the latitude and longitude of both the points of neutrino production and detection. On the other hand, for atmospheric neutrino experiment, shown in Fig.~\ref{Fig:Atmosphericaverage}, there are a whole host of neutrino velocities distributed on a two sphere. It is customary to bin events by zenith angle $\theta$, as shown in Fig.~\ref{Fig:Atmosphericaverage}. Then each bin corresponds to atmospheric neutrinos with the same path length from production to detection. For an experiment located near the earth's rotation axis, like IceCube \cite{IceCube:2017lak,IceCube:2017zcu}, this corresponds to averaging over $v_{||}$. However, for an experiment located elsewhere, like Super-Kamiokande \cite{Super-Kamiokande:2014exs,Super-Kamiokande:2014ndf,Super-Kamiokande:2017yvm}, the constant zenith angle averaging will correspond to different averages over the neutrino velocity depending on the particular location. To get an estimate of the impact of the velocity average on the vector DM effect, we can look at average over both neutrino velocity and DM polarization,  \bea\label{Eq:dailyangularaverage}
\langle P_{\nu_e\rarrow\nu_{\mu}}\rangle
&=&P^{(0)}+\frac{3}{8} \langle P^{(2)}\rangle+\cdots
\eea

The effect of daily average and the angular average is clear once we compare Eq.~\ref{Eq:phaseaverage} to Eq.~\ref{Eq:dailyangularaverage}.

\section{Experiments}
\label{Sec: Experiments}

In order to get a qualitative understanding of which experiments are the most efficient at detecting Vector DM, we can compare the relative strengths of various terms in the Hamiltonian

\bea
H_{\text{vac}} \sim \frac{\Delta m^2}{E_\nu}, 
\qquad H_{\text{int}} \sim g \frac{\sqrt{2\rho}}{\mA}.
\eea
Here, we can make a simplifying assumption that a given experiment is sensitive to DM when both terms in the Hamiltonian are comparable
\bea \label{Eq: bound estimate}
H_{int} \lesssim H_{vac} \qquad \to \qquad g \lesssim \frac{\mA}{E_{\nu}} \frac{\Delta m^2}{ \sqrt{2 \rho}} \sim 10^{-30} \l \frac{\mA}{10^{-20} \text{ eV}} \r \l \frac{10 \text{ GeV}}{E_{\nu}} \r.
\eea
As the vacuum term gets smaller with higher neutrino energies, we can see that experiments that operate at higher energies have greater discovery potential.

Naively, one could use the most energetic neutrinos to set the most stringent constraints. However, as the energy of neutrinos increases so does the oscillation length. Since the longest baseline we can probe is of the size of the earth, this limits the most efficient energies to $E \sim 10$ GeV. Experiments that operate or will operate at these energies are Super-Kamiokande \cite{Super-Kamiokande:2014exs,Super-Kamiokande:2014ndf,Super-Kamiokande:2017yvm}, IceCube \cite{IceCube:2017lak,IceCube:2017zcu}, DUNE \cite{DUNE:2015lol,DUNE:2020fgq,DUNE:2020jqi} and Hyper-Kamiokande \cite{Hyper-Kamiokande:2020aij}. We can divide these experiments into two categories: atmospheric neutrino experiments (Super-Kamiokande, IceCube, DUNE and Hyper-Kamiokande) and beam experiments (DUNE). Below, we consider one experiment of each type, with Super-Kamiokande as an example of an atmospheric neutrino experiment and DUNE as an example of a neutrino beam experiment. We focus on time-independent signatures of VDM, leaving the time-dependent analysis to future work. \footnote{By arguments around Eq. \eqref{Eq: bound estimate} we expect constraints from time-dependent and time-independent analysis to be comparable. It is important to note that the time-dependent analysis will be effective if the DM coherence time is at least as long as the duration of the experiment. }

\subsection{Super-Kamiokande}\label{subsec:SuperK}

Super-Kamiokande is a water-Cherenkov detector that is designed to measure neutrinos of astrophysical, atmospheric and accelerator origins. In this work, we focus on atmospheric neutrinos detected by Super-Kamiokande. In this type of experiment we can reconstruct neutrino direction and their energy. As neutrinos are produced in the atmosphere, the distance they travel before reaching the detector is related to the zenith angle $\theta$ shown in Fig.~\ref{Fig:Atmosphericaverage}. When $\cos \theta = 1$, the distance covered by neutrinos is of order $O(10\text{ km})$, while $\cos \theta = -1$ corresponds to distances of order $O(10^4\text{ km})$. In our analysis, we focus on neutrinos coming from underground ($\cos \theta <0$) as the longer effective baseline allows the effect of DM to accumulate, leading to stronger sensitivity.

Super-Kamiokande divides neutrino events into three main categories, fully contained (FC), partially contained (PC) and up-going muons (Up-$\mu$). Both FC and PC categories include neutrinos that convert into electrons/muons inside the detector. However, leptons produced during FC neutrino events deposit all energy inside the inner detector whereas leptons produced during PC events stop at the outer detector or deposit only a fraction of their energy before leaving the detector. The final type of event, Up-$\mu$, includes muons that were produced outside the detector but stopped inside it or passed through it leaving a trace. 

As we are interested in the neutrinos that have energies of order $10$ GeV and whose angular direction can be resolved, there are two event sub-categories that meet this criterion, PC through-going and Up-$\mu$ stopping. Both samples have similar energy distributions that peak around 10 GeV \cite{Super-Kamiokande:2014exs}. The first sub-category includes mostly muons and antimuons that were produced inside the detector but managed to escape it. These events are binned over zenith angle $\cos \theta \in [-1,1]$ with a bin size of $\Delta \cos \theta = 0.2$. The second sub-category contains the muons and antimuons that were produced outside the detector but stopped inside. Here, events are binned over a shorter range of zenith angle $\cos \theta \in [-1,0]$, with a smaller bin size of $\Delta \cos \theta = 0.1$.

In the analysis, we use two data sets for each sub-category. The first one contains the number of observed events \cite{Super-Kamiokande:2014ndf} as a function of the zenith angle $\cos \theta$, as shown in the Fig.~\ref{Fig:Atmosphericaverage}. The second data set has the ratio of observed neutrino events to the expectation in the case of standard neutrino oscillations as a function of zenith angle $\cos \theta$ \cite{Super-Kamiokande:2014exs}. For Up-$\mu$ stopping sub-category both data sets are binned over $\cos \theta \in [-1,0]$ into 10 bins of size $\Delta \cos \theta = 0.1$. For PC through-going sub-category we use the first 5 bins which correspond to $\cos \theta \in [-1,0]$ as they are the most sensitive to the effect of VDM.
We use these data sets as an experimental input to the $\chi^2$ test
\bea
\label{Eq: Chi Sq}
\chi^2(N_i,O_i) = 2 \sum\limits^{bins}_{i=1} \l N_i - O_i + O_i \ln \frac{O_i}{N_i} \r = 2 \sum\limits^{bins}_{i=1} O_{i} \l \l \frac{N_i}{N^{0}_i}x_{i}^{-1} \r - 1 + \ln x_i \frac{N^{0}_i}{N_i} \r
\eea
where $O_i$ is the number of observed events, $N_i$ is the number of expected events, $N^{0}_i$ is the number of expected events assuming standard neutrino oscillations \cite{Super-Kamiokande:2014exs} and $x_i = O_i/N^{0}_i$. The subscript $i$ refers to the bin.

The input from our model is contained in the ratio $\frac{N_i}{N^{0}_i}$. In our work, we make a simplifying assumption that the energy of neutrinos is $10$ GeV. We also assume that the detector response is identical for all neutrinos collected in a single bin. As a result, we can write the ratio $\frac{N_i}{N^{0}_i}$ as
\bea
\frac{N_i}{N^{0}_i} = \frac{ \sum\limits_{a = e, \mu} \bra P_{a \to \mu} \ket_{i} \Phi_{a,i} + \sum\limits_{a = \bar{e}, \bar{\mu}} \bra P_{ a \to  \bar{\mu}} \ket_{i} \Phi_{a,i}}
{\sum\limits_{a = e, \mu} \bra P^{0}_{a \to \mu} \ket_{i} \Phi_{a,i} + \sum\limits_{a = \bar{e}, \bar{\mu}} \bra P^{0}_{a \to \bar{\mu}} \ket_{i} \Phi_{a,i}}
\eea
where $\Phi_{a,i}$ is the atmospheric neutrino flux from \cite{Honda:2011nf}. $\bra P_{a \to \mu} \ket_{i}$ is the
probability averaged over bin $i$, where we average over azimuthal and zenith angles as well as DM oscillation and rotation of the earth. The probability is calculated in the low mass limit where we assume that the DM phase is approximately constant as the neutrino passes through the Earth. Following \cite{Super-Kamiokande:2014exs,Super-Kamiokande:2014ndf} we include the effect of changing electron density by employing the PREM model \cite{Dziewonski:1981xy}. Taking the same 3-flavor neutrino oscillation parameters as in \cite{Super-Kamiokande:2014exs} we get the following bounds at the $2 \sigma$ level
\bea
\label{Eq: Super-K Bound}
g< 2\cdot 10^{-31} \l \frac{\mA}{10^{-20} \text{ eV}}\r \quad \quad g'< 1\cdot 10^{-31} \l \frac{\mA}{10^{-20} \text{ eV}}\r.
\eea
These bounds are only valid in the low mass limit. The treatment of the high mass limit is explained in Sec. \ref{Sec: Results}. The projected bounds are shown in Fig.~\ref{fig: emu} and ~\ref{fig:mutau}.

In order to validate the above result we performed cross-checks by reproducing the results of Ref.~\cite{Super-Kamiokande:2014exs}.   Ref.~\cite{Super-Kamiokande:2014exs} examined the bounds coming from Lorentz symmetry violation in the form of chemical potentials and thus resembles our signal up to the time and direction dependence.
The reproduced results differed by up to a factor of two, which we can use as an estimate of the accuracy of our method. 

\subsection{DUNE}

As mentioned before, sensitivity is best for an experiment with a long baseline and high energy.  As an example of a neutrino beam experiment with these properties, we consider the Deep Underground Neutrino Experiment (DUNE).  DUNE produces a beam of primarily muon neutrinos at Fermilab at energies between $0.5$ and $10$ GeV. These neutrinos are then sent along a 1285 km path to a detector in Lead, SD. There, they measure the number of muon neutrinos that have disappeared ($N^{dis}$) and the number of electron neutrinos that have appeared ($N^{app}$).  Figures 10 and 11 in Ref.~\cite{DUNE:2020jqi} present a prediction for $N^{dis,app}$ using standard neutrino oscillations with the number of disappearance/appearance events. This data is presented in energy bins of width $0.25$ GeV from $0.5-8$ GeV. 

We can use these measurements to place bounds on our vector dark matter interaction. An exact treatment would involve a full simulation of the detector using the GLoBES framework~\cite{Huber:2004ka,Huber:2007ji} as was done in Ref.~\cite{DUNE:2020jqi} for standard oscillations and for a variety of BSM extensions to standard oscillations in Ref.~\cite{DUNE:2020fgq}. However, we take a simplified approach by assuming that the detector response is the same for all neutrinos in a single energy bin. In this approximation, we can write $N^{app}_i(N_i^{dis})$, the number of electron-neutrino appearance events (muon-neutrino disappearance events) in the $i^{th}$ energy bin, as
\bea
\label{Eq: Events Bin Integrals approx}
N^{dis}_i= N^{dis,0}_i\frac{1- \bra P_{\mu\rarrow \mu}\ket_i}{1-\bra P^0_{\mu\rarrow \mu}\ket_i} \qquad N^{app}_i=N^{app,0}_i\frac{\bra P_{\mu\rarrow e}\ket_i}{\bra P^0_{\mu\rarrow e}\ket_i}
\eea
Here the superscript $0$ represents quantities taken without dark matter interactions and $\bra\ket_i$ represents averaging over the $i^{th}$ energy bin. We numerically compute the oscillation probabilities using three flavor oscillations in the low mass limit. We vary the dark matter coupling, $g$ or $g'$, while  three-flavor oscillation parameters are kept fixed at the central values of the global fit~\cite{Esteban:2018azc} used by the DUNE collaboration in their simulated data for appearance/disappearance events~\cite{DUNE:2020fgq}.
Placing projected bounds from DUNE is complicated by the fact that we only have predicted data for $N^{app}_i$ and $N^{dis}_i$ and not any actual data. If we did have an observed number of events in each bin $O_i$ we could simply compute the $\chi^2$ for our theory using Eq ~\ref{Eq: Chi Sq}. If we wanted to compare our vector dark matter theory to standard oscillations, we would look at the difference in $\chi^2$ between the two theories
\bea
\label{Eq: Delta Chi Sq}
\Delta\chi^2(N_i,N_i^0,O_i)=\chi^2(N_i,O_i)-\chi^2(N^0_i,O_i)=2\sum_iN_i-N_i^0+O_i\ln\l\frac{N_i^0}{N_i}\r
\eea
Here the sum over bins sums over the bins for both muon-neutrino disappearance and electron-neutrino appearance to give the total $\chi^2$. However, since we do not have any real data to compare to, we imagine drawing $O_i$ from the distribution for appearance/disappearance events for standard oscillations given by $N_i^0$. Since $\Delta \chi^2$ is linear in $O_i$ this amounts to replacing $O_i$ with $N_i^0$ in Eq ~\ref{Eq: Delta Chi Sq}. 

\bea
\label{Eq: Delta Chi Sq avg}
\bra \Delta\chi^2(N_i,N_i^0,O_i)\ket_{O_i}=\chi^2(N_i,N^0_i)
\eea

If $\Delta\chi^2>4$ we would then be able to distinguish standard oscillations from standard oscillations with our vector dark matter interaction at the $2\sigma$ level. We can then place a predicted bound on our coupling, assuming that DUNE will not find disagreement with standard oscillations, by demanding 
\bea
\label{Eq: Bound Condition DUNE}
\Delta\chi^2(g)<4
\eea
where $g$ is the coupling of interest. As a cross-check, we were able to use this method to reproduce the bounds on the flavor-diagonal piece of the non-standard neutrino interactions from Ref.~\cite{DUNE:2020fgq} to within $20\%$. These flavor-diagonal non-standard neutrino interactions are very similar to our dark matter interaction. In fact, they are equivalent if we send $\cos(\mA t+\phi)\cos(\alpha)\rarrow 1$ in Eq.~\ref{Eq:intHamiltonian}. It is likely that our bounds have similar error bars.  

As seen from Eq.~\ref{Eq:intHamiltonian}, the strength of our interaction depends not only on the strength of the coupling $g$ and $g'$ but also on the relative direction of the neutrino velocity to the dark matter polarization and the phase of the dark matter.  All of these quantities vary on different timescales. The phase of the dark matter, $\mA t+\phi$ in Eq.~\ref{Eq:intHamiltonian}, changes on the time scale $\mA^{-1}$. The direction of the neutrinos' velocity changes on a daily time scale due to the earth's rotation. Finally, the direction and amplitude of the dark matter background and changes on the scale of the coherence time of the dark matter background which scales with $t_{coh}\sim 10^{6}\,\mA^{-1}$. For time scales that are shorter than the total data collection time of DUNE, around 3-7 years, we should average the probabilities $P_{\mu\rarrow\mu}$ and $P_{\mu \rarrow e}$ over these time scales before placing the bound from Eq.~\ref{Eq: Bound Condition DUNE}. This is true for all three of our relevant time scales except for the coherence time for masses $\mA<3 \cdot 10^{-18}$ eV. For masses below this limit, the dark photon field is coherent during the 3-7 year duration of the experiment so we must first compute bounds for each direction of the dark photon polarization. Then, in order to capture the fact we do not know in which direction the dark matter points, we should average these bounds over all dark photon polarizations. However, we find that whether or not we average over dark matter polarizations before or after computing the bounds leads to an $\mathcal{O}(10\%)$ change in the resulting average bound in the end. We, therefore, ignore this subtlety and average over all time scales before computing the bound. We find the bounds on $g(g')$ to be 
\bea
\label{Eq: DUNE Bounds}
g< 10^{-30} \l \frac{\mA}{10^{-20} \text{ eV}}\r \quad \quad g'< 7\cdot 10^{-31} \l \frac{\mA}{10^{-20} \text{ eV}}\r .
\eea
These bounds are shown plotted below in Fig.~\ref{fig: emu} and ~\ref{fig:mutau}. 

As mentioned before, we are using DUNE as an example of the sensitivity of a neutrino beam experiment.  DUNE expects to detect oscillations from the background atmospheric neutrinos as described in Ref.~\cite{DUNE:2015lol} where they argue DUNE will have a sensitivity similar to Hyper-Kamiokande.  If one were to use this aspect of DUNE, then the bounds in Eq.~\ref{Eq: DUNE Bounds} and Eq.~\ref{Eq: Super-K Bound} would be improved upon significantly.

\subsection{Results}
\label{Sec: Results}

\begin{figure}[t]
    \centering
    \includegraphics[width=0.75\textwidth]{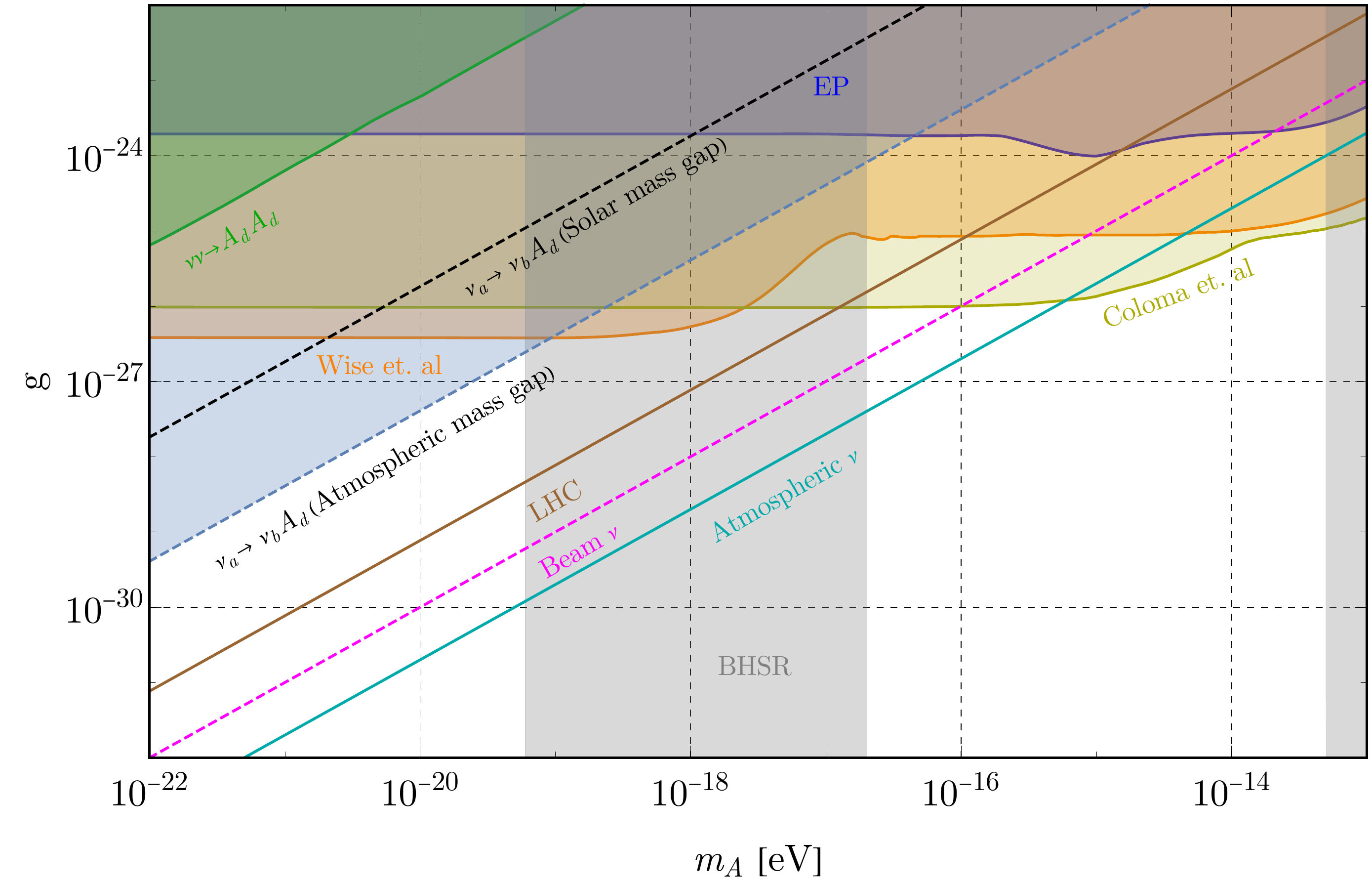}
    \caption{Constraints on the $L_e-L_{\mu}$ coupling from atmospheric and beam neutrino experiments. The curves show upper limits on the coupling $g$. The atmospheric neutrino bound is represented by the cyan line, which is derived using Super-Kamiokande data \cite{Super-Kamiokande:2014exs,Super-Kamiokande:2014ndf}. The dashed magenta line gives the projected sensitivity of neutrino beam experiments, which is obtained assuming DUNE operating in the beam mode \cite{DUNE:2020fgq}.  Our constraints are several orders of magnitude stronger than the cosmological $\nu$-decay bound (light blue and black dashed lines)~\cite{Chen:2022idm} and five orders of magnitude stronger than other terrestrial bounds at the lower end of the allowed mass range. It is expected that time dependent analysis will improve these bounds. Other constraints are from BH superradiance (gray band)~\cite{Baryakhtar:2017ngi}, neutrino oscillations modified by a fifth force (orange and yellow lines)~\cite{Wise:2018rnb,Coloma:2020gfv}, $\Delta N_{\text{eff}}$ through $\nu\nu\rightarrow \Ad\Ad$ (green line)~\cite{Dror:2020fbh,Huang:2017egl}, EP violating forces (dark blue line)~\cite{Wise:2018rnb,Schlamminger:2007ht} and a model dependent bound coming from a mono-lepton plus missing energy search at LHC (brown line)~\cite{Ekhterachian:2021rkx}.
    }
    \label{fig: emu}
\end{figure}

We numerically find the DUNE and Super-Kamiokande bound for a DM mass in the small mass limit using the methods described before.  To obtain a bound for an arbitrary mass, we rescale the bound using the full second-order correction to the oscillation probability, $\bra P^{(2)}\ket$, given in Eq.~\ref{Eq:smallbaseline}. In Eq.~\ref{Eq:smallbaseline}, $L$ is the neutrino baseline, $\Delta_\nu=\frac{\Delta m_{13}^2}{4E_\nu}$, $A_0$ is the field strength $A_0=\frac{\sqrt{2\rho_{DM}}}{\mA}$, and $g$ is the coupling we are interested in (either $g$ or $g'$). For each of the experiments, we insert different values of $L$ and $E_\nu$. For DUNE~\cite{DUNE:2020jqi}, we use the baseline $L\approx 1300$ km and $E_\nu=2$ GeV, roughly the energy at which detection events peak. For Super-Kamiokande we use $E_\nu =10$ GeV. The length for Super-Kamiokande is complicated since, for atmospheric neutrinos, $L$ varies depending on the zenith angle $\theta$. In principle, we should use an effective length, weighted over the bins. For simplicity, we use $L\approx 12000$ km, roughly the diameter of the earth. A different choice of $L$ does not alter our constraint in the low mass regime or its qualitative feature over the whole of the parameter space.

In order to use Eq.~\ref{Eq:smallbaseline} to rescale our bounds, perturbation theory in $g$ ($g'$) must be valid.  Using Eq.~\ref{Eq: Perturbative limit} with $E_\nu=10$ GeV for Super-Kamiokande and $E_\nu=2$ GeV for DUNE to obtain the perturbative limit for both experiments and comparing these limits to the bounds placed in Eq.~\ref{Eq: Super-K Bound} and ~\ref{Eq: DUNE Bounds}, we find that our bounds are soundly in the perturbative limit.

\begin{figure}[t]
     \centering
    \includegraphics[width=0.75\textwidth]{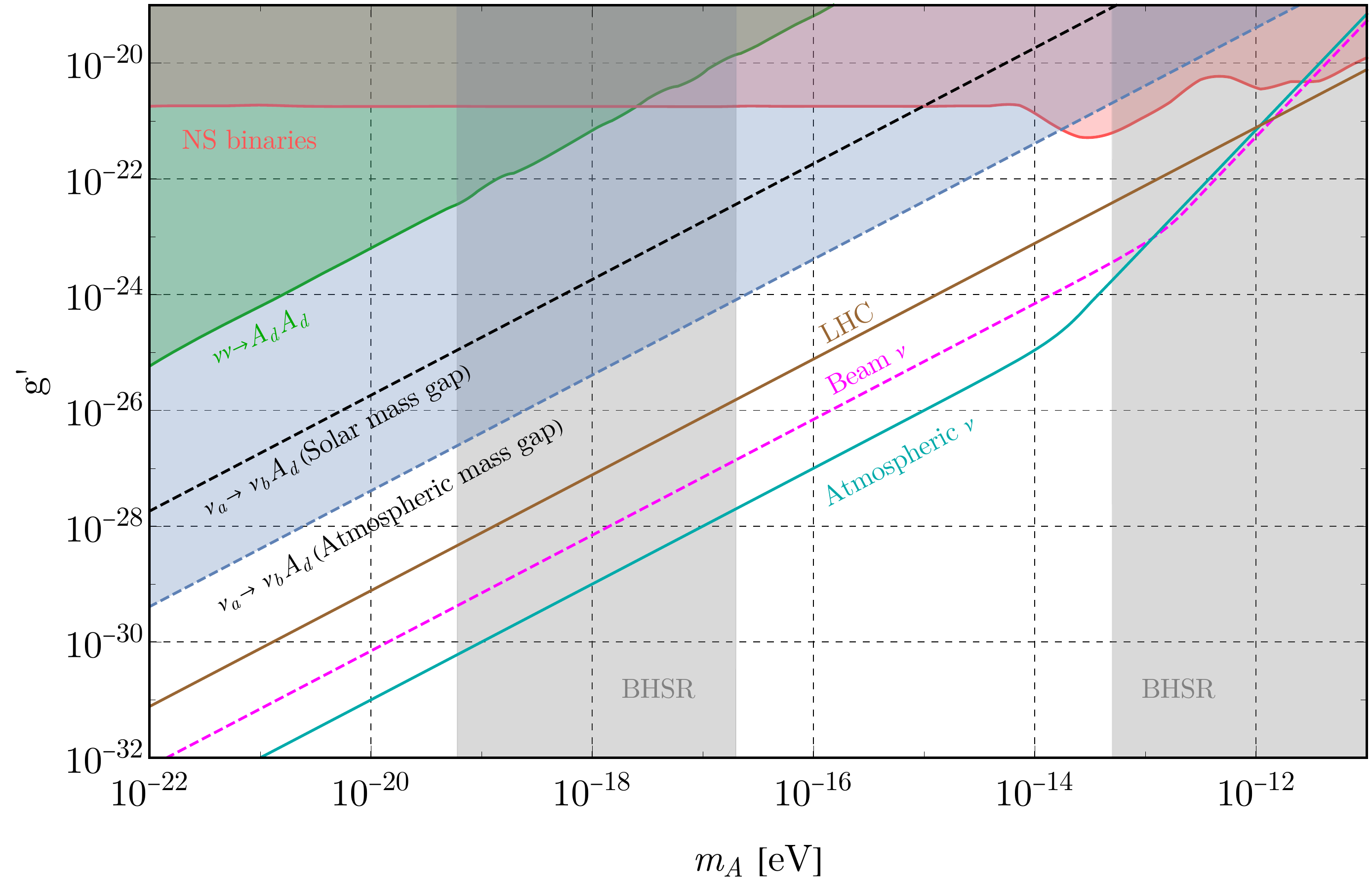}
    \caption{Constraints on the $L_\mu-L_{\tau}$ coupling from atmospheric and beam neutrino experiments. The curves show upper limits on the coupling $g'$. The atmospheric neutrino bound is derived using Super-Kamiokande data \cite{Super-Kamiokande:2014exs,Super-Kamiokande:2014ndf}. The projected sensitivity of beam neutrino experiments is obtained assuming DUNE operating in the beam mode \cite{DUNE:2020fgq}. Our constraints are several orders of magnitude stronger than the cosmological $\nu$-decay bound (light blue and black dashed lines)~\cite{Chen:2022idm} and eight orders of magnitude stronger than other existing constraints at the lower end of the allowed mass range. It is expected that time-dependent analysis will improve these bounds. Previous constraints are from BH superradiance (gray band)~\cite{Baryakhtar:2017ngi}, $\Delta N_{\text{eff}}$ through $\nu\nu\rightarrow \Ad\Ad$ (green line)~\cite{Dror:2020fbh,Huang:2017egl}, NS binaries (red line)~\cite{Dror:2019uea} and a model-dependent bound coming from a mono-lepton plus missing energy search at LHC (brown line)~\cite{Ekhterachian:2021rkx}. 
    }
    \label{fig:mutau}
\end{figure}

As an additional cross-check of this extrapolation, we solved for the transition probabilities numerically for a few values of $\mA$ in the high mass regime for both DUNE and Super-Kamiokande. While doing this for all points in our parameters space is too computationally expensive, we find good agreement between our perturbation extrapolation to high masses and the exact numerical results.

\section{Conclusions}
\label{Sec: Conclusions}

In this work, we studied the effects of ultra-light vector dark matter consisting of a dark photon. We consider our dark photon to be the gauge boson of gauged lepton flavor number with different couplings for each flavor. We showed the effect of such a coupling in the context of neutrino oscillations is to give each neutrino flavor an effective chemical potential proportional to the background dark matter field and the coupling. Because the couplings are flavor asymmetric, these chemical potentials are different for each neutrino flavor and thus will have some effect on neutrino oscillations. Because this interaction is diagonal in flavor space, we saw that the net effect of these couplings is to dampen the mixing, and thus dampen oscillations between neutrino flavor eigenstates. 

Using this effect, we are able to place bounds on the flavor asymmetric couplings $g$ ($U(1)_{L_{e}-L_{\mu}}$) and $g'$ ($U(1)_{L_{\mu}-L_{\tau}}$) by comparing to neutrino oscillation experiments.  Using Super-Kamiokande (DUNE) as an example, we found the (projected) bounds coming from an atmospheric (beam) neutrino experiment.  These bounds are shown in Fig.~\ref{fig: emu} and Fig.~\ref{fig:mutau}. For both, we found our bounds beat previously leading bounds for low dark matter masses and give bounds comparable to leading bounds at high masses. 

We also considered time-dependent effects unique to our vector dark matter background. In particular, due to the Earth's rotation, neutrinos in the earth's frame see a vector background that rotates on a daily basis. Since the coupling between neutrinos and the dark matter background is proportional to $\Vec{A}_d \cdot \Vec{v}_{\nu}$, this rotation effect would manifest as daily modulations in neutrino oscillations. In our bounds, we averaged over these daily modulations. However, one could look for these daily modulations directly. These daily modulation would be seen as a daily time dependence in the appearance/disappearance events. One would need to have the time-stamped data to see this effect. This would involve working directly with the data from neutrino oscillation experiments and so we leave it for future works. 
In addition to this time-dependent analysis, bounds could be made more precise with future atmospheric neutrino experiments such as Hyper-Kamiokande, DUNE and by working more directly with experiments like IceCube.

\section*{Acknowledgments}

We thank Aaron Vincent for useful discussions. DB, SD, AH and CR are supported in part by the NSF under Grant 
No. PHY-2210361 and by the Maryland Center for Fundamental Physics 
(MCFP). This research was supported in part by Perimeter Institute for Theoretical Physics. Research at Perimeter
Institute is supported by the Government of Canada through the Department of Innovation, Science and
Economic Development and by the Province of Ontario through the Ministry of Research and Innovation. 

\bibliography{main}

\providecommand{\href}[2]{#2}\begingroup\raggedright\begin{thebibliography}{10}

\bibitem{Bergstrom:2000pn}
L.~Bergstr\"om, {\it {Nonbaryonic dark matter: Observational evidence and
  detection methods}},  {\em Rept. Prog. Phys.} {\bf 63} (2000) 793,
  [\href{http://arxiv.org/abs/hep-ph/0002126}{{\tt hep-ph/0002126}}].

\bibitem{Bertone:2004pz}
G.~Bertone, D.~Hooper, and J.~Silk, {\it {Particle dark matter: Evidence,
  candidates and constraints}},  {\em Phys. Rept.} {\bf 405} (2005) 279--390,
  [\href{http://arxiv.org/abs/hep-ph/0404175}{{\tt hep-ph/0404175}}].

\bibitem{Lisanti:2016jxe}
M.~Lisanti, {\it {Lectures on Dark Matter Physics}},  in {\em {Theoretical
  Advanced Study Institute in Elementary Particle Physics}: {New Frontiers in
  Fields and Strings}}, pp.~399--446, 2017.
\newblock \href{http://arxiv.org/abs/1603.03797}{{\tt arXiv:1603.03797}}.

\bibitem{Arvanitaki:2009fg}
A.~Arvanitaki, S.~Dimopoulos, S.~Dubovsky, N.~Kaloper, and J.~March-Russell,
  {\it {String Axiverse}},  {\em Phys. Rev. D} {\bf 81} (2010) 123530,
  [\href{http://arxiv.org/abs/0905.4720}{{\tt arXiv:0905.4720}}].

\bibitem{Hui:2016ltb}
L.~Hui, J.~P. Ostriker, S.~Tremaine, and E.~Witten, {\it {Ultralight scalars as
  cosmological dark matter}},  {\em Phys. Rev. D} {\bf 95} (2017), no.~4
  043541, [\href{http://arxiv.org/abs/1610.08297}{{\tt arXiv:1610.08297}}].

\bibitem{Hu:2000ke}
W.~Hu, R.~Barkana, and A.~Gruzinov, {\it {Cold and fuzzy dark matter}},  {\em
  Phys. Rev. Lett.} {\bf 85} (2000) 1158--1161,
  [\href{http://arxiv.org/abs/astro-ph/0003365}{{\tt astro-ph/0003365}}].

\bibitem{Irsic:2017yje}
V.~Ir\v{s}i\v{c}, M.~Viel, M.~G. Haehnelt, J.~S. Bolton, and G.~D. Becker, {\it
  {First constraints on fuzzy dark matter from Lyman-$\alpha$ forest data and
  hydrodynamical simulations}},  {\em Phys. Rev. Lett.} {\bf 119} (2017), no.~3
  031302, [\href{http://arxiv.org/abs/1703.04683}{{\tt arXiv:1703.04683}}].

\bibitem{Dentler:2021zij}
M.~Dentler, D.~J.~E. Marsh, R.~Hlo\v{z}ek, A.~Lagu\"e, K.~K. Rogers, and
  D.~Grin, {\it {Fuzzy dark matter and the Dark Energy Survey Year 1 data}},
  {\em Mon. Not. Roy. Astron. Soc.} {\bf 515} (2022), no.~4 5646--5664,
  [\href{http://arxiv.org/abs/2111.01199}{{\tt arXiv:2111.01199}}].

\bibitem{Hui:2021tkt}
L.~Hui, {\it {Wave Dark Matter}},  {\em Ann. Rev. Astron. Astrophys.} {\bf 59}
  (2021) 247--289, [\href{http://arxiv.org/abs/2101.11735}{{\tt
  arXiv:2101.11735}}].

\bibitem{Khmelnitsky:2013lxt}
A.~Khmelnitsky and V.~Rubakov, {\it {Pulsar timing signal from ultralight
  scalar dark matter}},  {\em JCAP} {\bf 02} (2014) 019,
  [\href{http://arxiv.org/abs/1309.5888}{{\tt arXiv:1309.5888}}].

\bibitem{Graham:2015ouw}
P.~W. Graham, I.~G. Irastorza, S.~K. Lamoreaux, A.~Lindner, and K.~A. van
  Bibber, {\it {Experimental Searches for the Axion and Axion-Like Particles}},
   {\em Ann. Rev. Nucl. Part. Sci.} {\bf 65} (2015) 485--514,
  [\href{http://arxiv.org/abs/1602.00039}{{\tt arXiv:1602.00039}}].

\bibitem{Graham:2015ifn}
P.~W. Graham, D.~E. Kaplan, J.~Mardon, S.~Rajendran, and W.~A. Terrano, {\it
  {Dark Matter Direct Detection with Accelerometers}},  {\em Phys. Rev. D} {\bf
  93} (2016), no.~7 075029, [\href{http://arxiv.org/abs/1512.06165}{{\tt
  arXiv:1512.06165}}].

\bibitem{Arvanitaki:2014faa}
A.~Arvanitaki, J.~Huang, and K.~Van~Tilburg, {\it {Searching for dilaton dark
  matter with atomic clocks}},  {\em Phys. Rev. D} {\bf 91} (2015), no.~1
  015015, [\href{http://arxiv.org/abs/1405.2925}{{\tt arXiv:1405.2925}}].

\bibitem{Geraci:2018fax}
A.~A. Geraci, C.~Bradley, D.~Gao, J.~Weinstein, and A.~Derevianko, {\it
  {Searching for Ultralight Dark Matter with Optical Cavities}},  {\em Phys.
  Rev. Lett.} {\bf 123} (2019), no.~3 031304,
  [\href{http://arxiv.org/abs/1808.00540}{{\tt arXiv:1808.00540}}].

\bibitem{Irastorza:2018dyq}
I.~G. Irastorza and J.~Redondo, {\it {New experimental approaches in the search
  for axion-like particles}},  {\em Prog. Part. Nucl. Phys.} {\bf 102} (2018)
  89--159, [\href{http://arxiv.org/abs/1801.08127}{{\tt arXiv:1801.08127}}].

\bibitem{PhysRevA.93.063630}
Y.~V. Stadnik and V.~V. Flambaum, {\it Enhanced effects of variation of the
  fundamental constants in laser interferometers and application to dark-matter
  detection},  {\em Phys. Rev. A} {\bf 93} (Jun, 2016) 063630.

\bibitem{Hook:2018jle}
A.~Hook, {\it {Solving the Hierarchy Problem Discretely}},  {\em Phys. Rev.
  Lett.} {\bf 120} (2018), no.~26 261802,
  [\href{http://arxiv.org/abs/1802.10093}{{\tt arXiv:1802.10093}}].

\bibitem{DiLuzio:2021gos}
L.~Di~Luzio, B.~Gavela, P.~Quilez, and A.~Ringwald, {\it {Dark matter from an
  even lighter QCD axion: trapped misalignment}},  {\em JCAP} {\bf 10} (2021)
  001, [\href{http://arxiv.org/abs/2102.01082}{{\tt arXiv:2102.01082}}].

\bibitem{DiLuzio:2021pxd}
L.~Di~Luzio, B.~Gavela, P.~Quilez, and A.~Ringwald, {\it {An even lighter QCD
  axion}},  {\em JHEP} {\bf 05} (2021) 184,
  [\href{http://arxiv.org/abs/2102.00012}{{\tt arXiv:2102.00012}}].

\bibitem{Graham:2013gfa}
P.~W. Graham and S.~Rajendran, {\it {New Observables for Direct Detection of
  Axion Dark Matter}},  {\em Phys. Rev. D} {\bf 88} (2013) 035023,
  [\href{http://arxiv.org/abs/1306.6088}{{\tt arXiv:1306.6088}}].

\bibitem{Budker:2013hfa}
D.~Budker, P.~W. Graham, M.~Ledbetter, S.~Rajendran, and A.~Sushkov, {\it
  {Proposal for a Cosmic Axion Spin Precession Experiment (CASPEr)}},  {\em
  Phys. Rev. X} {\bf 4} (2014), no.~2 021030,
  [\href{http://arxiv.org/abs/1306.6089}{{\tt arXiv:1306.6089}}].

\bibitem{ADMX:2010ubl}
{\bf ADMX} Collaboration, A.~Wagner et~al., {\it {A Search for Hidden Sector
  Photons with ADMX}},  {\em Phys. Rev. Lett.} {\bf 105} (2010) 171801,
  [\href{http://arxiv.org/abs/1007.3766}{{\tt arXiv:1007.3766}}].

\bibitem{Chaudhuri:2014dla}
S.~Chaudhuri, P.~W. Graham, K.~Irwin, J.~Mardon, S.~Rajendran, and Y.~Zhao,
  {\it {Radio for hidden-photon dark matter detection}},  {\em Phys. Rev. D}
  {\bf 92} (2015), no.~7 075012, [\href{http://arxiv.org/abs/1411.7382}{{\tt
  arXiv:1411.7382}}].

\bibitem{Knirck:2018ojz}
S.~Knirck, T.~Yamazaki, Y.~Okesaku, S.~Asai, T.~Idehara, and T.~Inada, {\it
  {First results from a hidden photon dark matter search in the meV sector
  using a plane-parabolic mirror system}},  {\em JCAP} {\bf 11} (2018) 031,
  [\href{http://arxiv.org/abs/1806.05120}{{\tt arXiv:1806.05120}}].

\bibitem{An:2014twa}
H.~An, M.~Pospelov, J.~Pradler, and A.~Ritz, {\it {Direct Detection Constraints
  on Dark Photon Dark Matter}},  {\em Phys. Lett. B} {\bf 747} (2015) 331--338,
  [\href{http://arxiv.org/abs/1412.8378}{{\tt arXiv:1412.8378}}].

\bibitem{Bloch:2016sjj}
I.~M. Bloch, R.~Essig, K.~Tobioka, T.~Volansky, and T.-T. Yu, {\it {Searching
  for Dark Absorption with Direct Detection Experiments}},  {\em JHEP} {\bf 06}
  (2017) 087, [\href{http://arxiv.org/abs/1608.02123}{{\tt arXiv:1608.02123}}].

\bibitem{Carney:2019cio}
D.~Carney, A.~Hook, Z.~Liu, J.~M. Taylor, and Y.~Zhao, {\it {Ultralight dark
  matter detection with mechanical quantum sensors}},  {\em New J. Phys.} {\bf
  23} (2021), no.~2 023041, [\href{http://arxiv.org/abs/1908.04797}{{\tt
  arXiv:1908.04797}}].

\bibitem{Pierce:2018xmy}
A.~Pierce, K.~Riles, and Y.~Zhao, {\it {Searching for Dark Photon Dark Matter
  with Gravitational Wave Detectors}},  {\em Phys. Rev. Lett.} {\bf 121}
  (2018), no.~6 061102, [\href{http://arxiv.org/abs/1801.10161}{{\tt
  arXiv:1801.10161}}].

\bibitem{Dror:2019uea}
J.~A. Dror, R.~Laha, and T.~Opferkuch, {\it {Probing muonic forces with neutron
  star binaries}},  {\em Phys. Rev. D} {\bf 102} (2020), no.~2 023005,
  [\href{http://arxiv.org/abs/1909.12845}{{\tt arXiv:1909.12845}}].

\bibitem{Fabbrichesi:2020wbt}
M.~Fabbrichesi, E.~Gabrielli, and G.~Lanfranchi, {\it {The Dark Photon}},
  \href{http://arxiv.org/abs/2005.01515}{{\tt arXiv:2005.01515}}.

\bibitem{Dev:2020kgz}
A.~Dev, P.~A.~N. Machado, and P.~Mart\'\i{}nez-Mirav\'e, {\it {Signatures of
  ultralight dark matter in neutrino oscillation experiments}},  {\em JHEP}
  {\bf 01} (2021) 094, [\href{http://arxiv.org/abs/2007.03590}{{\tt
  arXiv:2007.03590}}].

\bibitem{Reynoso:2016hjr}
M.~M. Reynoso and O.~A. Sampayo, {\it {Propagation of high-energy neutrinos in
  a background of ultralight scalar dark matter}},  {\em Astropart. Phys.} {\bf
  82} (2016) 10--20, [\href{http://arxiv.org/abs/1605.09671}{{\tt
  arXiv:1605.09671}}].

\bibitem{Berlin:2016woy}
A.~Berlin, {\it {Neutrino Oscillations as a Probe of Light Scalar Dark
  Matter}},  {\em Phys. Rev. Lett.} {\bf 117} (2016), no.~23 231801,
  [\href{http://arxiv.org/abs/1608.01307}{{\tt arXiv:1608.01307}}].

\bibitem{Krnjaic:2017zlz}
G.~Krnjaic, P.~A.~N. Machado, and L.~Necib, {\it {Distorted neutrino
  oscillations from time varying cosmic fields}},  {\em Phys. Rev. D} {\bf 97}
  (2018), no.~7 075017, [\href{http://arxiv.org/abs/1705.06740}{{\tt
  arXiv:1705.06740}}].

\bibitem{Brdar:2017kbt}
V.~Brdar, J.~Kopp, J.~Liu, P.~Prass, and X.-P. Wang, {\it {Fuzzy dark matter
  and nonstandard neutrino interactions}},  {\em Phys. Rev. D} {\bf 97} (2018),
  no.~4 043001, [\href{http://arxiv.org/abs/1705.09455}{{\tt
  arXiv:1705.09455}}].

\bibitem{Davoudiasl:2018hjw}
H.~Davoudiasl, G.~Mohlabeng, and M.~Sullivan, {\it {Galactic Dark Matter
  Population as the Source of Neutrino Masses}},  {\em Phys. Rev. D} {\bf 98}
  (2018), no.~2 021301, [\href{http://arxiv.org/abs/1803.00012}{{\tt
  arXiv:1803.00012}}].

\bibitem{Liao:2018byh}
J.~Liao, D.~Marfatia, and K.~Whisnant, {\it {Light scalar dark matter at
  neutrino oscillation experiments}},  {\em JHEP} {\bf 04} (2018) 136,
  [\href{http://arxiv.org/abs/1803.01773}{{\tt arXiv:1803.01773}}].

\bibitem{Capozzi:2018bps}
F.~Capozzi, I.~M. Shoemaker, and L.~Vecchi, {\it {Neutrino Oscillations in Dark
  Backgrounds}},  {\em JCAP} {\bf 07} (2018) 004,
  [\href{http://arxiv.org/abs/1804.05117}{{\tt arXiv:1804.05117}}].

\bibitem{Huang:2018cwo}
G.-Y. Huang and N.~Nath, {\it {Neutrinophilic Axion-Like Dark Matter}},  {\em
  Eur. Phys. J. C} {\bf 78} (2018), no.~11 922,
  [\href{http://arxiv.org/abs/1809.01111}{{\tt arXiv:1809.01111}}].

\bibitem{Farzan:2019yvo}
Y.~Farzan, {\it {Ultra-light scalar saving the 3 + 1 neutrino scheme from the
  cosmological bounds}},  {\em Phys. Lett. B} {\bf 797} (2019) 134911,
  [\href{http://arxiv.org/abs/1907.04271}{{\tt arXiv:1907.04271}}].

\bibitem{Cline:2019seo}
J.~M. Cline, {\it {Viable secret neutrino interactions with ultralight dark
  matter}},  {\em Phys. Lett. B} {\bf 802} (2020) 135182,
  [\href{http://arxiv.org/abs/1908.02278}{{\tt arXiv:1908.02278}}].

\bibitem{Losada:2021bxx}
M.~Losada, Y.~Nir, G.~Perez, and Y.~Shpilman, {\it {Probing scalar dark matter
  oscillations with neutrino oscillations}},  {\em JHEP} {\bf 04} (2022) 030,
  [\href{http://arxiv.org/abs/2107.10865}{{\tt arXiv:2107.10865}}].

\bibitem{Huang:2021kam}
G.-y. Huang and N.~Nath, {\it {Neutrino meets ultralight dark matter:
  0\ensuremath{\nu}\ensuremath{\beta}\ensuremath{\beta} decay and cosmology}},
  {\em JCAP} {\bf 05} (2022), no.~05 034,
  [\href{http://arxiv.org/abs/2111.08732}{{\tt arXiv:2111.08732}}].

\bibitem{Chun:2021ief}
E.~J. Chun, {\it {Neutrino Transition in Dark Matter}},
  \href{http://arxiv.org/abs/2112.05057}{{\tt arXiv:2112.05057}}.

\bibitem{Losada:2022uvr}
M.~Losada, Y.~Nir, G.~Perez, I.~Savoray, and Y.~Shpilman, {\it {Parametric
  resonance in neutrino oscillations induced by ultra-light dark matter and
  implications for KamLAND and JUNO}},
  \href{http://arxiv.org/abs/2205.09769}{{\tt arXiv:2205.09769}}.

\bibitem{Dev:2022bae}
A.~Dev, G.~Krnjaic, P.~Machado, and H.~Ramani, {\it {Constraining Feeble
  Neutrino Interactions with Ultralight Dark Matter}},
  \href{http://arxiv.org/abs/2205.06821}{{\tt arXiv:2205.06821}}.

\bibitem{Wolfenstein:1977ue}
L.~Wolfenstein, {\it {Neutrino Oscillations in Matter}},  {\em Phys. Rev. D}
  {\bf 17} (1978) 2369--2374.

\bibitem{Mikheyev:1985zog}
S.~P. Mikheyev and A.~Y. Smirnov, {\it {Resonance Amplification of Oscillations
  in Matter and Spectroscopy of Solar Neutrinos}},  {\em Sov. J. Nucl. Phys.}
  {\bf 42} (1985) 913--917.

\bibitem{Super-Kamiokande:2017yvm}
{\bf Super-Kamiokande} Collaboration, K.~Abe et~al., {\it {Atmospheric neutrino
  oscillation analysis with external constraints in Super-Kamiokande I-IV}},
  {\em Phys. Rev. D} {\bf 97} (2018), no.~7 072001,
  [\href{http://arxiv.org/abs/1710.09126}{{\tt arXiv:1710.09126}}].

\bibitem{Alonso-Alvarez:2021pgy}
G.~Alonso-\'Alvarez and J.~M. Cline, {\it {Sterile neutrino dark matter
  catalyzed by a very light dark photon}},  {\em JCAP} {\bf 10} (2021) 041,
  [\href{http://arxiv.org/abs/2107.07524}{{\tt arXiv:2107.07524}}].

\bibitem{DUNE:2015lol}
{\bf DUNE} Collaboration, R.~Acciarri et~al., {\it {Long-Baseline Neutrino
  Facility (LBNF) and Deep Underground Neutrino Experiment (DUNE)}: {Conceptual
  Design Report, Volume 2: The Physics Program for DUNE at LBNF}},
  \href{http://arxiv.org/abs/1512.06148}{{\tt arXiv:1512.06148}}.

\bibitem{DUNE:2020fgq}
{\bf DUNE} Collaboration, B.~Abi et~al., {\it {Prospects for beyond the
  Standard Model physics searches at the Deep Underground Neutrino
  Experiment}},  {\em Eur. Phys. J. C} {\bf 81} (2021), no.~4 322,
  [\href{http://arxiv.org/abs/2008.12769}{{\tt arXiv:2008.12769}}].

\bibitem{DUNE:2020jqi}
{\bf DUNE} Collaboration, B.~Abi et~al., {\it {Long-baseline neutrino
  oscillation physics potential of the DUNE experiment}},  {\em Eur. Phys. J.
  C} {\bf 80} (2020), no.~10 978, [\href{http://arxiv.org/abs/2006.16043}{{\tt
  arXiv:2006.16043}}].

\bibitem{IceCube:2017lak}
{\bf IceCube} Collaboration, M.~G. Aartsen et~al., {\it {Measurement of
  Atmospheric Neutrino Oscillations at 6\textendash{}56 GeV with IceCube
  DeepCore}},  {\em Phys. Rev. Lett.} {\bf 120} (2018), no.~7 071801,
  [\href{http://arxiv.org/abs/1707.07081}{{\tt arXiv:1707.07081}}].

\bibitem{IceCube:2017zcu}
{\bf IceCube} Collaboration, M.~G. Aartsen et~al., {\it {Search for Nonstandard
  Neutrino Interactions with IceCube DeepCore}},  {\em Phys. Rev. D} {\bf 97}
  (2018), no.~7 072009, [\href{http://arxiv.org/abs/1709.07079}{{\tt
  arXiv:1709.07079}}].

\bibitem{Super-Kamiokande:2014exs}
{\bf Super-Kamiokande} Collaboration, K.~Abe et~al., {\it {Test of Lorentz
  invariance with atmospheric neutrinos}},  {\em Phys. Rev. D} {\bf 91} (2015),
  no.~5 052003, [\href{http://arxiv.org/abs/1410.4267}{{\tt arXiv:1410.4267}}].

\bibitem{Super-Kamiokande:2014ndf}
{\bf Super-Kamiokande} Collaboration, K.~Abe et~al., {\it {Limits on sterile
  neutrino mixing using atmospheric neutrinos in Super-Kamiokande}},  {\em
  Phys. Rev. D} {\bf 91} (2015) 052019,
  [\href{http://arxiv.org/abs/1410.2008}{{\tt arXiv:1410.2008}}].

\bibitem{Hyper-Kamiokande:2020aij}
{\bf Hyper-Kamiokande} Collaboration, K.~Abe et~al., {\it {The Hyper-Kamiokande
  Experiment - Snowmass LOI}},  \href{http://arxiv.org/abs/2009.00794}{{\tt
  arXiv:2009.00794}}.

\bibitem{Honda:2011nf}
M.~Honda, T.~Kajita, K.~Kasahara, and S.~Midorikawa, {\it {Improvement of low
  energy atmospheric neutrino flux calculation using the JAM nuclear
  interaction model}},  {\em Phys. Rev. D} {\bf 83} (2011) 123001,
  [\href{http://arxiv.org/abs/1102.2688}{{\tt arXiv:1102.2688}}].

\bibitem{Dziewonski:1981xy}
A.~M. Dziewonski and D.~L. Anderson, {\it {Preliminary reference earth model}},
   {\em Phys. Earth Planet. Interiors} {\bf 25} (1981) 297--356.

\bibitem{Huber:2004ka}
P.~Huber, M.~Lindner, and W.~Winter, {\it {Simulation of long-baseline neutrino
  oscillation experiments with GLoBES (General Long Baseline Experiment
  Simulator)}},  {\em Comput. Phys. Commun.} {\bf 167} (2005) 195,
  [\href{http://arxiv.org/abs/hep-ph/0407333}{{\tt hep-ph/0407333}}].

\bibitem{Huber:2007ji}
P.~Huber, J.~Kopp, M.~Lindner, M.~Rolinec, and W.~Winter, {\it {New features in
  the simulation of neutrino oscillation experiments with GLoBES 3.0: General
  Long Baseline Experiment Simulator}},  {\em Comput. Phys. Commun.} {\bf 177}
  (2007) 432--438, [\href{http://arxiv.org/abs/hep-ph/0701187}{{\tt
  hep-ph/0701187}}].

\bibitem{Esteban:2018azc}
I.~Esteban, M.~C. Gonzalez-Garcia, A.~Hernandez-Cabezudo, M.~Maltoni, and
  T.~Schwetz, {\it {Global analysis of three-flavour neutrino oscillations:
  synergies and tensions in the determination of $\theta_{23}$, $\delta_{CP}$,
  and the mass ordering}},  {\em JHEP} {\bf 01} (2019) 106,
  [\href{http://arxiv.org/abs/1811.05487}{{\tt arXiv:1811.05487}}].

\bibitem{Chen:2022idm}
J.~Z. Chen, I.~M. Oldengott, G.~Pierobon, and Y.~Y.~Y. Wong, {\it {Weaker yet
  again: mass spectrum-consistent cosmological constraints on the neutrino
  lifetime}},  {\em Eur. Phys. J. C} {\bf 82} (2022), no.~7 640,
  [\href{http://arxiv.org/abs/2203.09075}{{\tt arXiv:2203.09075}}].

\bibitem{Baryakhtar:2017ngi}
M.~Baryakhtar, R.~Lasenby, and M.~Teo, {\it {Black Hole Superradiance
  Signatures of Ultralight Vectors}},  {\em Phys. Rev. D} {\bf 96} (2017),
  no.~3 035019, [\href{http://arxiv.org/abs/1704.05081}{{\tt
  arXiv:1704.05081}}].

\bibitem{Wise:2018rnb}
M.~B. Wise and Y.~Zhang, {\it {Lepton Flavorful Fifth Force and Depth-dependent
  Neutrino Matter Interactions}},  {\em JHEP} {\bf 06} (2018) 053,
  [\href{http://arxiv.org/abs/1803.00591}{{\tt arXiv:1803.00591}}].

\bibitem{Coloma:2020gfv}
P.~Coloma, M.~C. Gonzalez-Garcia, and M.~Maltoni, {\it {Neutrino oscillation
  constraints on U(1)' models: from non-standard interactions to long-range
  forces}},  {\em JHEP} {\bf 01} (2021) 114,
  [\href{http://arxiv.org/abs/2009.14220}{{\tt arXiv:2009.14220}}]. [Erratum:
  JHEP 11, 115 (2022)].

\bibitem{Dror:2020fbh}
J.~A. Dror, {\it {Discovering leptonic forces using nonconserved currents}},
  {\em Phys. Rev. D} {\bf 101} (2020), no.~9 095013,
  [\href{http://arxiv.org/abs/2004.04750}{{\tt arXiv:2004.04750}}].

\bibitem{Huang:2017egl}
G.-y. Huang, T.~Ohlsson, and S.~Zhou, {\it {Observational Constraints on Secret
  Neutrino Interactions from Big Bang Nucleosynthesis}},  {\em Phys. Rev. D}
  {\bf 97} (2018), no.~7 075009, [\href{http://arxiv.org/abs/1712.04792}{{\tt
  arXiv:1712.04792}}].

\bibitem{Schlamminger:2007ht}
S.~Schlamminger, K.~Y. Choi, T.~A. Wagner, J.~H. Gundlach, and E.~G.
  Adelberger, {\it {Test of the equivalence principle using a rotating torsion
  balance}},  {\em Phys. Rev. Lett.} {\bf 100} (2008) 041101,
  [\href{http://arxiv.org/abs/0712.0607}{{\tt arXiv:0712.0607}}].

\bibitem{Ekhterachian:2021rkx}
M.~Ekhterachian, A.~Hook, S.~Kumar, and Y.~Tsai, {\it {Bounds on gauge bosons
  coupled to nonconserved currents}},  {\em Phys. Rev. D} {\bf 104} (2021),
  no.~3 035034, [\href{http://arxiv.org/abs/2103.13396}{{\tt
  arXiv:2103.13396}}].

\end{thebibliography}\endgroup
\bibliographystyle{JHEP}
\end{document}